\documentclass[preprint2]{aastex}

\usepackage{natbib}

\slugcomment{Accepted by PASP}

\shorttitle{Hi-GAL: the Herschel Galactic Plane Survey}
\shortauthors{Molinari et al.}

\newcommand{\higal}{Hi-GAL}

\newcommand{\subsun}{$_{\odot}$}
\newcommand{\um}{$\mu$m}
\newcommand{\adeg}{$^{\circ}$}
\newcommand{\hi}{H{\sc i}}
\newcommand{\hii}{H{\sc ii}}
\newcommand{\msun}{M$_{\odot}$}

\begin{document}


\title{Hi-GAL: the {\sc {\textit {Herschel}}} infrared Galactic Plane Survey}


\author{S. Molinari\altaffilmark{1,2}, B. Swinyard\altaffilmark{44},  J. Bally\altaffilmark{6},
  M. Barlow\altaffilmark{8}, J.-P. Bernard\altaffilmark{9},
  P. Martin\altaffilmark{31}, T. Moore\altaffilmark{46},
  A. Noriega-Crespo\altaffilmark{19}, R. Plume\altaffilmark{48},
  L. Testi\altaffilmark{21,55}
% ...
}

\altaffiltext{1}{PI of the Hi-GAL project, \email{sergio.molinari@ifsi-roma.inaf.it}}
\altaffiltext{2}{INAF-IFSI, Fosso del Cavaliere 100, 00133 Roma, Italy}
% ...
\altaffiltext{6}{Center for Astrophysics and Space Astronomy (CASA),
  Department of Astrophysical and Planetary Sciences, University of
  Colorado, Boulder, USA}
% ...
\altaffiltext{8}{University College London, UK}
\altaffiltext{9}{Centre d'Etude Spatiale du Rayonnement, Toulouse, France}
% ...
\altaffiltext{19}{Spitzer Science Center, California Institute of
  Technology, USA}
% ...
\altaffiltext{21}{INAF - Osservatorio Astrofisico di Arcetri, Italy}
% ...
\altaffiltext{31}{University of Toronto, CITA, Canada}
% ...
\altaffiltext{44}{STFC, Rutherford Appleton Laboratory, UK}
% ...
\altaffiltext{46}{Astrophysics Research Institute, Liverpool John
  Moores University, UK}
% ...
\altaffiltext{48}{Department of Physics \& Astronomy, University of
  Calgary, Canada}
% ...
\altaffiltext{55}{European Southern Observatory, Garching, Germany}
% ...





\begin{abstract}

\higal, the Herschel infrared Galactic Plane Survey, is an Open Time
Key Project of the Herschel Space Observatory.  It will make an
unbiased photometric  
survey of the inner Galactic Plane by mapping a two-degree wide strip
in the longitude range $\mid l \mid < 60^{\circ}$  
in five wavebands between 70$\mu$m and 500$\mu$m.  The aim of
\higal\ is to detect the earliest phases of the formation  
of molecular clouds and high-mass stars and to use the optimum
combination of Herschel wavelength coverage,  
sensitivity, mapping strategy and speed to deliver a homogeneous
census of star-forming regions and  
cold structures in the interstellar medium. The resulting
representative samples will yield the variation  
of source temperature, luminosity, mass and age in a wide range of
Galactic environments at all scales from  
massive YSOs in protoclusters to entire spiral arms, providing an
evolutionary sequence for the formation of intermediate and high-mass
stars.  This information is essential to the formulation of a  
predictive global model of the role of environment and feedback in
regulating the star-formation process.  
Such a model is vital to understanding star formation on galactic
scales and in the early Universe.  Hi-GAL  
will also provide a science legacy for decades to come with
incalculable potential for systematic and 
serendipitous science in a wide range of astronomical fields, enabling
the optimum use of future major  
facilities such as JWST and ALMA.

\end{abstract}


\section{Introduction}

Dust is the most robust tracer of the `Galactic
ecology' - the cycling of material from dying stars to the ionized,
atomic, and molecular phases of the ISM, into star forming cloud cores,
and back into stars. While atoms, ions, and molecules are imperfect
tracers because they undergo complex phase changes, chemical processing,
depletion onto grains, and are subject to complex excitation
conditions, dust is relatively stable in most phases of the ISM. It is optically
thin in the Far Infrared (FIR) over most of the Galaxy, so that its
emission and absorption simply depend on emissivity, column density and
temperature. Cold dust in particular (10\,K$\le$T$\le 40$\,K) traces the
bulk of non-stellar baryonic mass in all of the above ``habitats" of the
Galactic ecosystem.

Temperature and luminosity and, as their by-product, mass of cold dust measured
over the entire Galactic Plane (GP), are, at sub-parsec resolution, the
critical quantities needed to formulate a global predictive model of
the cycling process between the Galactic ISM and star formation.  This
process drives the galactic ecology in normal spirals as well as the enhanced
star-formation rates of starburst galaxies and mergers and a quantitative 
understanding of it is needed in order to follow the formation and evolution 
of galaxies throughout the cosmos. The adequate measurement of these key 
quantities has been beyond
the capabilities of the previous mid- to far-infrared surveys of the
Galactic Plane (IRAS, \citealt{neu84}; MSX, \citealt{pri01}; COBE/DIRBE and FIRAS, e.g. \citealt{sod94}; ISO, \citealt{omo03}; Spitzer, \citealt{ben03, car09}) either due to limited wavelength coverage and/or inadequate
spatial resolution leading to confusion. The balloon-borne BLAST experiment \citep{pasc08} implements Herschel/SPIRE detector arrays and is providing exciting anticipations of what Herschel will do. The AKARI satellite \citep{mur07} improves over IRAS, and results from its FIR photometric mapping of the GP are eagerly awaited.

Observing the distribution and temperature of dust across the Galaxy
will resolve many current debates  such as the modes of
formation of molecular clouds and high-mass stars. 

Molecular clouds are traditionally thought to follow a ``slow
formation" scheme, where distributed 
material is accumulated by  large-scale perturbations such as the
passage of a spiral arm. Shielding by dust and surface reactions on
grains promotes  the \hi $\rightarrow$H$_2$ transition, which in turn
allows the formation of other molecules that cool the cloud. Gravity,
mediated by magnetic fields, leads to star formation. In this
scenario cloud lifetimes are about $\sim$30 Myr  (\citealt{lei89}).
This picture has difficulty explaining the 
absence of quiescent,  non star forming GMCs (however, see
\citealt{pal97}) and the continuous re-generation of turbulence 
needed to support GMCs for many crossing times.  Alternatively, a
``fast formation" scenario has been proposed 
(\citealt{har01}) in which most MCs are transient,
short-lived structures (\citealt{sto98,pad99}) created in the
post-shock regions of converging large-scale flows.  Stars form on
very short timescales (\citealt{elm00}).  However, rapid
MC formation requires rapid \hi $\rightarrow$H$_2$ conversion
(\citealt{gol05}). Accelerated H$_2$ formation requires either 
high-density pre-shock conditions (n$\sim$200 cm$^{-3}$, T$\leq$100K;
\citealt{pri01}), or strong turbulence (\citealt{glo07}), higher than
observed.   

On the other hand, the formation of high-mass 
stars and of the star clusters hosting them is likely the most
important process that shapes the formation and evolution of
galaxies. Massive stars are responsible for the global ionization of the ISM. 
Their energetic stellar winds and supernova blast waves direct the
dynamical evolution of the ISM, shaping its morphology, energetics and
chemistry, and influencing the formation of subsequent generations of
stars and planetary systems. Despite their importance, remarkably little is known about how 
massive stars form (\citealt{mck03}).
We lack a ``fundamental theory" or, rather, a galaxy-scale
predictive model for star formation. One of the main limitations
to this goal is the lack of statistically significant and well-characterized samples of young massive stars in the various evolutionary stages and environments on which a theory can be based. In turn, this results from the difficulty of gathering observational data on
on a large number of forming high-mass stars: they make up only a very small fraction 
of the total number of stars in the Galaxy, their early evolutionary  
phases of massive stars are more rapid than those of low-mass stars,
they lie at large distance and form in crowded environments. 
 
\begin{figure*}[t!]
\begin{tabular}{lr}
\includegraphics[width=8cm]{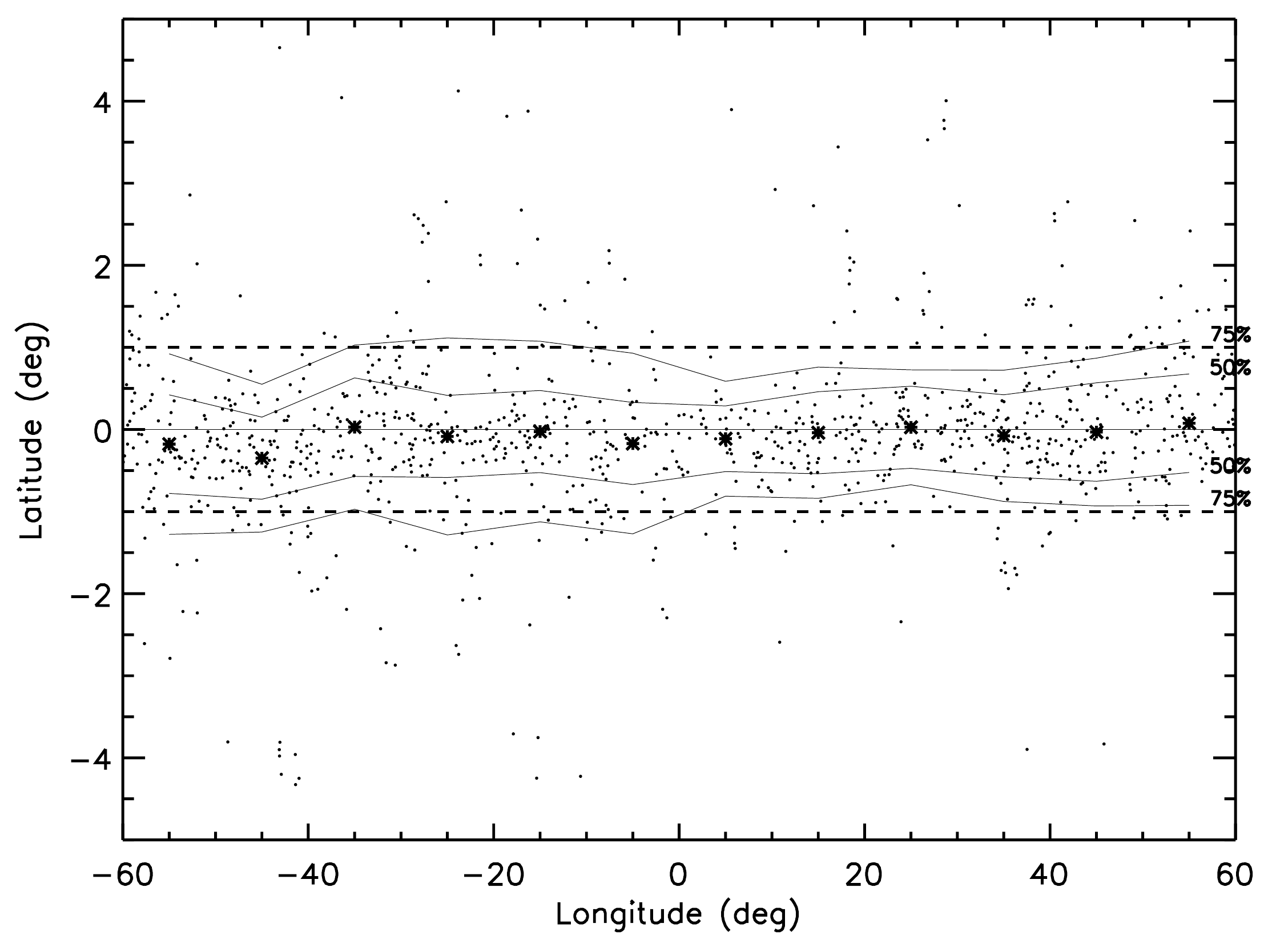} &
\includegraphics[width=8cm]{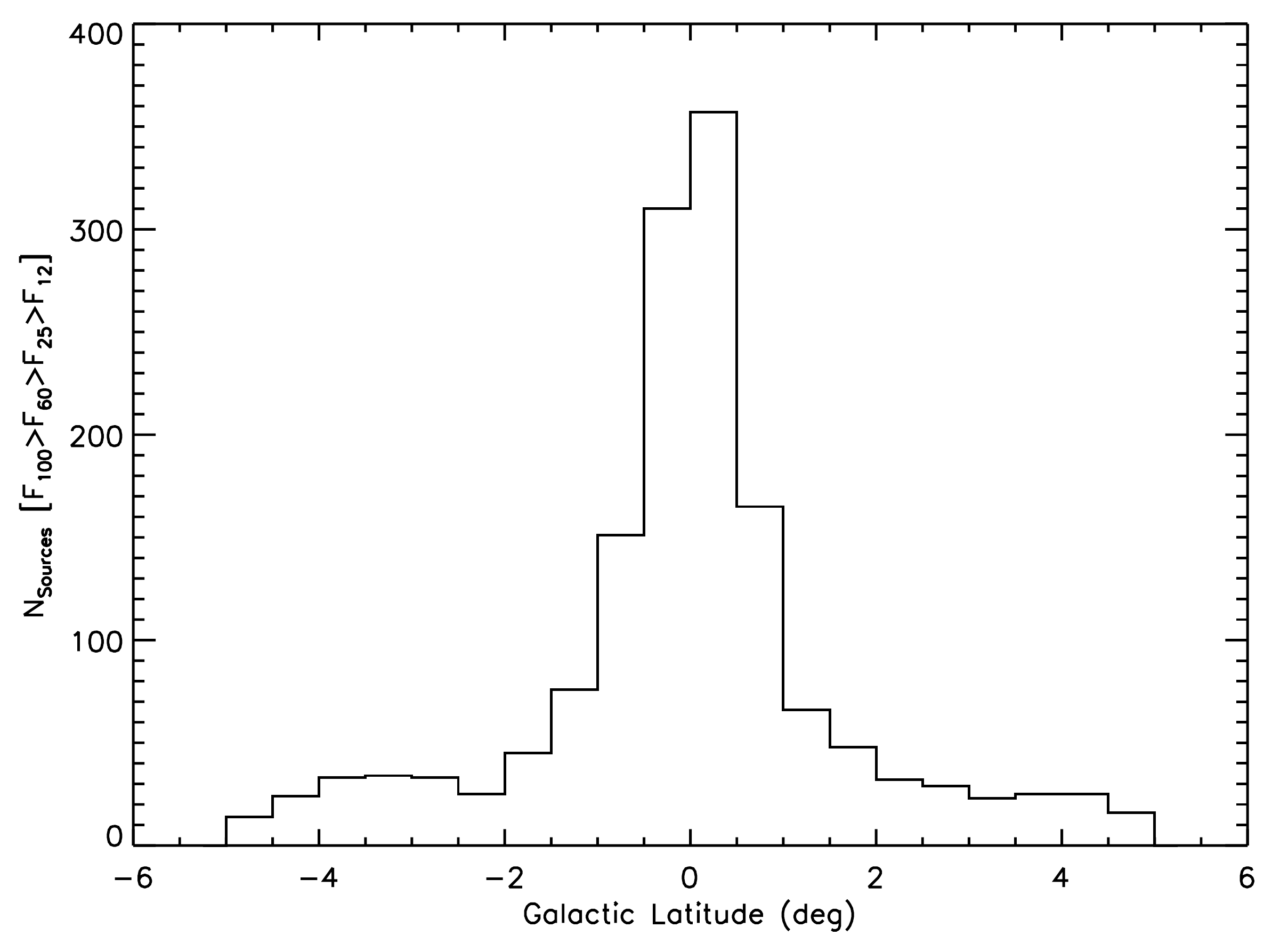} 
\end{tabular}
\caption{Left panel: $l-b$ plot of $\lambda$-rising SED IRAS sources; the straight thin line is the Galactic
  midplane. The asterisks mark the median latitude of the sources
  computed in 10\adeg $l$ bins. The thin lines delimit the
  regions where 50\% and 75\% of $|b| \leq 5$\adeg\ IRAS sources are
  contained. Right panel: $b$-distribution of the same
  IRAS sources in the $|l|\leq 60$\adeg\ region.\label{source-distr}} 
\end{figure*}

There is thus a long list of questions that the community has
been addressing for some time, not finding satisfactory answers. Here is 
an abridged list:
\begin{list}{\labelitemi}{\leftmargin=1em}
\item  What is the temperature and density structure of the ISM?  How do
  molecular clouds form, evolve, and how are they disrupted? 
\item What is the origin of the stellar initial mass function (IMF)?
  What is its relationship to the mass function (MF) of ISM structures
  and cloud cores on all scales?  
\item  How do massive stars and clusters form and how do they evolve?  What
are the earliest stages of massive star formation and what are the timescales
of these early phases?
\item  How do the Star Formation Rate (SFR) and Efficiency (SFE)
  vary as a function of Galactocentric distance and environmental
  conditions such as the intensity of the Interstellar Radiation Field
  (ISRF), ISM metallicity, proximity to spiral arms or the molecular
  ring, external triggers, and total pressure?  
\item Does a threshold column density for star formation exist in our 
Galaxy? What determines the value of this possible threshold? 
\item What are the physical processes involved in triggered star formation
on all scales and how does triggered star formation differ from
spontaneous star formation?
\item How do the local properties of the ISM and the rates of
  spontaneous or triggered star formation relate to the global scaling
  laws observed in external galaxies ?  
\end{list}

Using the Herschel telescope, the largest ever in space, Hi-GAL, the Herschel infrared Galactic
Plane survey, will provide unique new data with which to address these questions.  Hi-GAL will make  thermal infrared maps of the Galactic Plane at a spatial resolution 30 times better than IRAS and 100 times better than DIRBE, from which a complete census of compact source luminosities, masses, and spectral energy distributions (SEDs) will be derived. Source distances are a crucial parameter in this respect, and a dedicated effort will be needed (see \S\ref{othersurveys}). Extraction of statistically significant samples of star-forming regions
and cold ISM structures will be possible in all the environments of
the Milky Way at all scales from massive Young
Stellar Objects (YSOs) in individual protoclusters
to complete spiral arms.

In the following we present the specific characteristics of
the survey as well as some of the science outcomes that we expect to obtain
with this unique project.

\section{\higal\ Observing Strategy}

The area covered by \higal\ ($\mid l \mid \leq 
60^{\circ}, \mid b \mid \leq 
1^{\circ}$) contains most of the star formation in the Galaxy, and it is the one
which offers the best coverage in ancillary data  which will be
critical in the scientific analysis (see \S\ref{othersurveys}). The
$b$ distribution and extent of the survey is shown in
Fig.~\ref{source-distr} along with the $l-b$ plot of $\lambda$-rising SED IRAS sources  ($F_{100} > F_{60} > F_{25} > F_{12}$) which are potential 
YSOs. The \higal\ area (thick dashed lines in that figure) represents the
$|b|\leq$1\adeg\ strip 
centered on the midplane and contains about 80\% of all potential YSOs
contained in $|b|\leq 5$\adeg\ strip, thus encompassing most of the
potential star formation sites in the inner Galaxy.

The Herschel photometric cameras PACS (\citealt{pogetal08}) and SPIRE
(\citealt{gri09}) will be used in parallel mode (pMode\footnote{In
  pMode the Herschel telescope is scanning the 
sky in a raster fashion at constant speed while both PACS and SPIRE
acquire data simultaneously}) to maximize survey speed 
and wavelength coverage. Due to the instruments wavelength multiplexing
capabilities, one pMode observation delivers maps at five different
wavelengths: 70 and 170\um\ with PACS and 250, 350 and 500\um\ with
SPIRE. Both cameras cameras use bolometric detector arrays to map the
sky by scanning the spacecraft along approximate great circles. Both
instruments require their on board sub-kelvin coolers to be recycled
to provide the detectors with the operating temperature required of 
about 0.3 K in each case. In pMode both instruments are placed into
their photometric observing mode 
with the detectors at their correct operating temperature, i.e. both
instrument coolers are recycled, and data are taken from the five arrays
simultaneously as the spacecraft is scanned across the sky. 

\begin{figure}[h]
\resizebox{\hsize}{!}{\includegraphics[width=10.3cm]{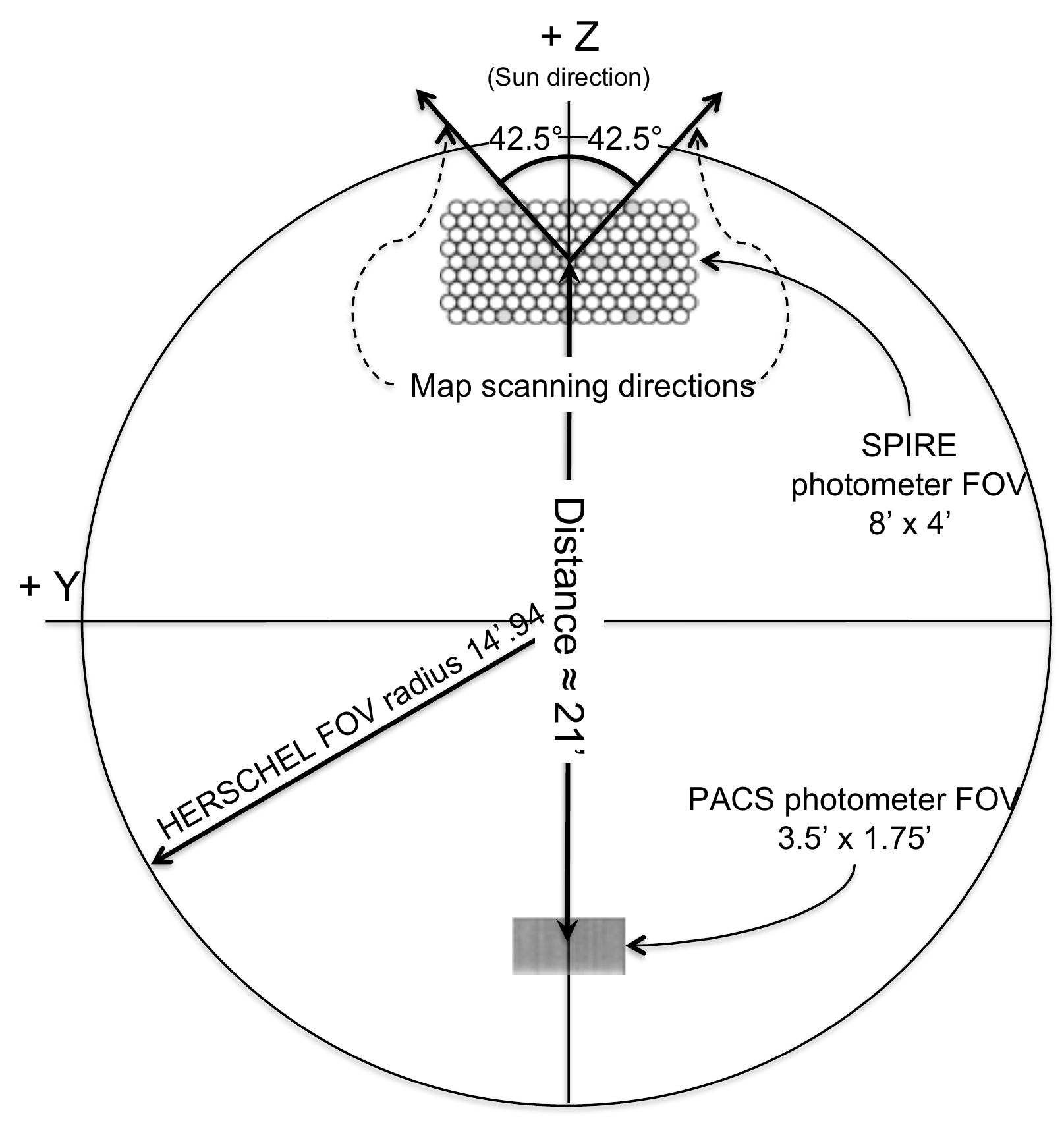}}
\caption{The field of view of the PACS and SPIRE instruments shown in
  the context of the Herschel field of view as viewed in the
  co-ordinate system of the spacecraft +Z refers to the axis towards
  the Sun.  The +X axis is along the telescope boresight out of the
  page.  The scan directions used to map the sky are as indicated. The different photometric channels of each instrument map the same region of the sky.}  
\label{focalplane}
\end{figure}

\begin{figure}[h!]
\resizebox{\hsize}{!}{\includegraphics[width=8.3cm]{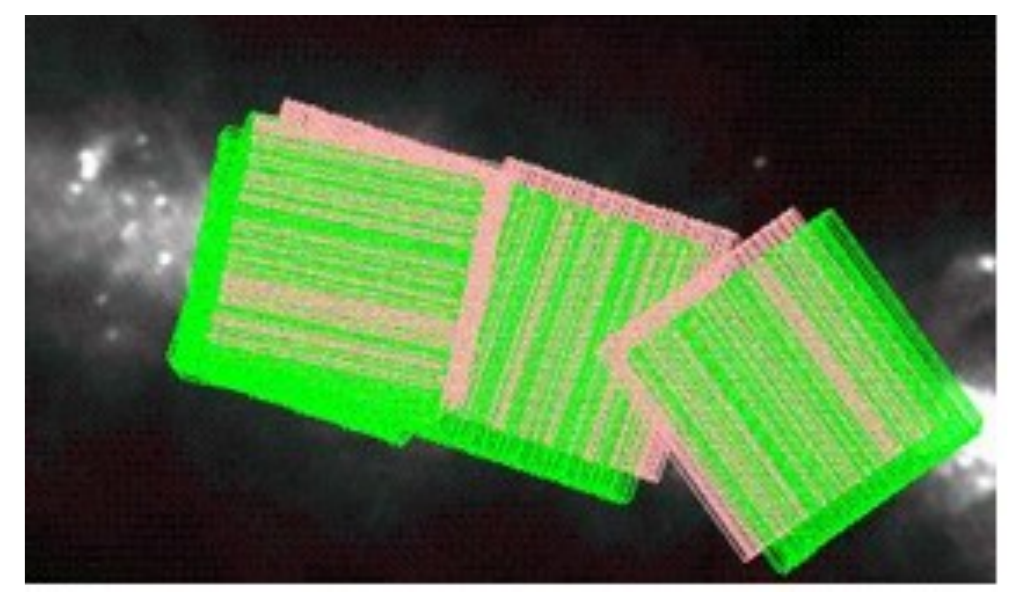}}
\caption{Sample AORs (Astronomical Observation Request) overlaid on the IRAS 100\um\ image of a portion of the
  Galactic plane.  From left to right we
  show: both nominal and orthogonal 2.2\adeg x2.2\adeg\ pMode AORs
  overlaid on one another. The PACS-covered area is outlined in pink
  while SPIRE is green.}  
\label{hspot_example}
\end{figure}

The size and separation of the fields of view of PACS and SPIRE are
shown in Fig.~\ref{focalplane} as viewed in the spacecraft co-ordinate system.
Although the PACS array fully spatially samples the point
spread function from the telescope it still has gaps between the
sub-arrays, and the SPIRE arrays only sparsely sample the sky.  In
order to make fully spatially sampled maps it is necessary to scan the
SPIRE array at an angle of ~42.5$^{\circ}$ with respect to its short
symmetry axis.  Scanning at an angle is also used for the PACS arrays
to fill in for the gaps between sub-arrays.  To achieve redundancy in
the data and remove instrumental effects such as high-frequency
detector response, slow drifts in gain or stray light, saturation and
environmental (cirrus confusion) effects it is also necessary to make
at least a second pass over the same region of the sky using the
other scan angle at $-42.5$\adeg\ angle which, quite conveniently, is nearly orthogonal to the first one (see fig.~\ref{hspot_example}).

The distance between each scan in parallel mode is set by the size of
the PACS array (being the smaller of the two), and the effective length of each leg of the raster takes into account the separation between the two fields
of view.  HSPOT, the Herschel-SPOT observing tool\footnote{ftp://ftp.sciops.esa.int/pub/hspot/HSpot\_download.html}, 
automatically 
calculates these parameters to ensure that the area required is
covered. The distance between scans is approximately 155\arcsec\ and
the excess length of the scan beyond the required length to cover the
area is typically 20\arcmin .  An example of how the sky is covered in
a Parallel Mode observation used in \higal\ is shown in
Fig.~\ref{hspot_example}.

The strategy employed to cover the $-60$\adeg$\leq\ l \leq$ 60\adeg,
$|b|\leq$1\adeg\ survey area is to conduct observations with series of
55 2.2\adeg\ $\times 2.2$\adeg\ square  tiles with two passes over each
tile with the two above mentioned scan angles (see Fig. \ref{focalplane}).  These tiles will be spaced every 2\adeg, so that the overlap between tiles ensures that no coverage gaps are introduced by different tile orientation due to variable  satellite roll angles with time; fig.~\ref{hspot_example} shows a section of the galactic plane with
consecutive observing blocks overlaid providing overlapping coverage.

Given the
spatial separation required for PACS in the pMode 
observations, the SPIRE data is heavily oversampled and we cover a
greater area than required with each individual instrument than would
be required using them sequentially.   Although it might seem that
sequential PACS and SPIRE scan mode observations would be more
efficient in fact the satellite overheads, set up, calibration and
pointing acquisition, {\it et cetera},  mean that it requires 30\% more time to cover
the same area sequentially compared to using the pMode.   

In order to cover the maximum area in the shortest time \higal\ data
will be taken at the maximum possible scan speed for the satellite of 60
\arcsec /sec.  This implies a beam crossing time for the short
wavelength, 250 \um\ band of SPIRE of ~3\,Hz  well within the
bandwidth available in the detectors of 5 Hz.  However, although the PACS detectors have a similar response time the much smaller point
spread function will be smeared out compared to that achievable with a
slower scan.  Additionally because of  the finite data transmission
bandwidth between the Herschel satellite and the ground, it is
necessary to perform on-board data compression for the PACS data which
are the most demanding in terms of number of pixels (2048 for the
70\um\ array and 512 for the 170\um\ array) at the frame acquisition
rate of 40Hz. The baseline configuration for the pMode is then
to average on board groups of 8 frames at 70\um\ and 4 frames at
170\um. Since the telescope is continuously scanning while acquiring,
this coaddition will result in a further degradation of the Point
Spread Function in the direction of the scan from its original
diffraction limited shape; the effect will be more severe at 70\um\ where the degradation should be of a factor two based on simulations. This loss in imaging fidelity at the shortest wavelength is considered
acceptable for a survey like Hi-GAL because, as discussed in the
previous section, our main focus is toward a large-scale picture of
the galaxy. Taking advantage of the orthogonal cross scan observing strategy, we may be able to
recover some of the spatial resolution by careful deconvolution during
post processing. 

\subsection{Detection of compact sources}

The SPIRE digital readout electronics impose a limitation on the
brightest sources that can be observed for a given offset setting (DC
voltage removal) before digitization (SPIRE Instrument Users Manual,
2007).  This problem can be alleviated to some extent by choosing a
bias setting that gives the largest dynamic range per offset range.
Simulations of the effect of bias variation show that setting a bias
higher ($\sim$3x) than the predicted nominal value will approximately double
the dynamic range for most detectors under the conditions likely to be
found in orbit (telescope temperature and emissivity and sky
background).  The same simulations show that a significant ($>$10\%) fraction 
of the SPIRE 250\um\ array detectors will saturate on sources greater
than 500 Jy. The situation is slightly more relaxed for the 350 and
500\um\ arrays.  We take the upper limit of detectable sources in the
SPIRE bands as 500 Jy/beam assuming that a strong source instrument
setting is used.  This setting is required for all observations of
bright regions/sources with SPIRE and is not a special Hi-GAL
configuration.  The saturation limits for PACS should be around 2000Jy
at nominal bias, that will be used for the Hi-GAL survey. 

\begin{figure}[h!]
\resizebox{\hsize}{!}{\includegraphics[width=8.3cm]{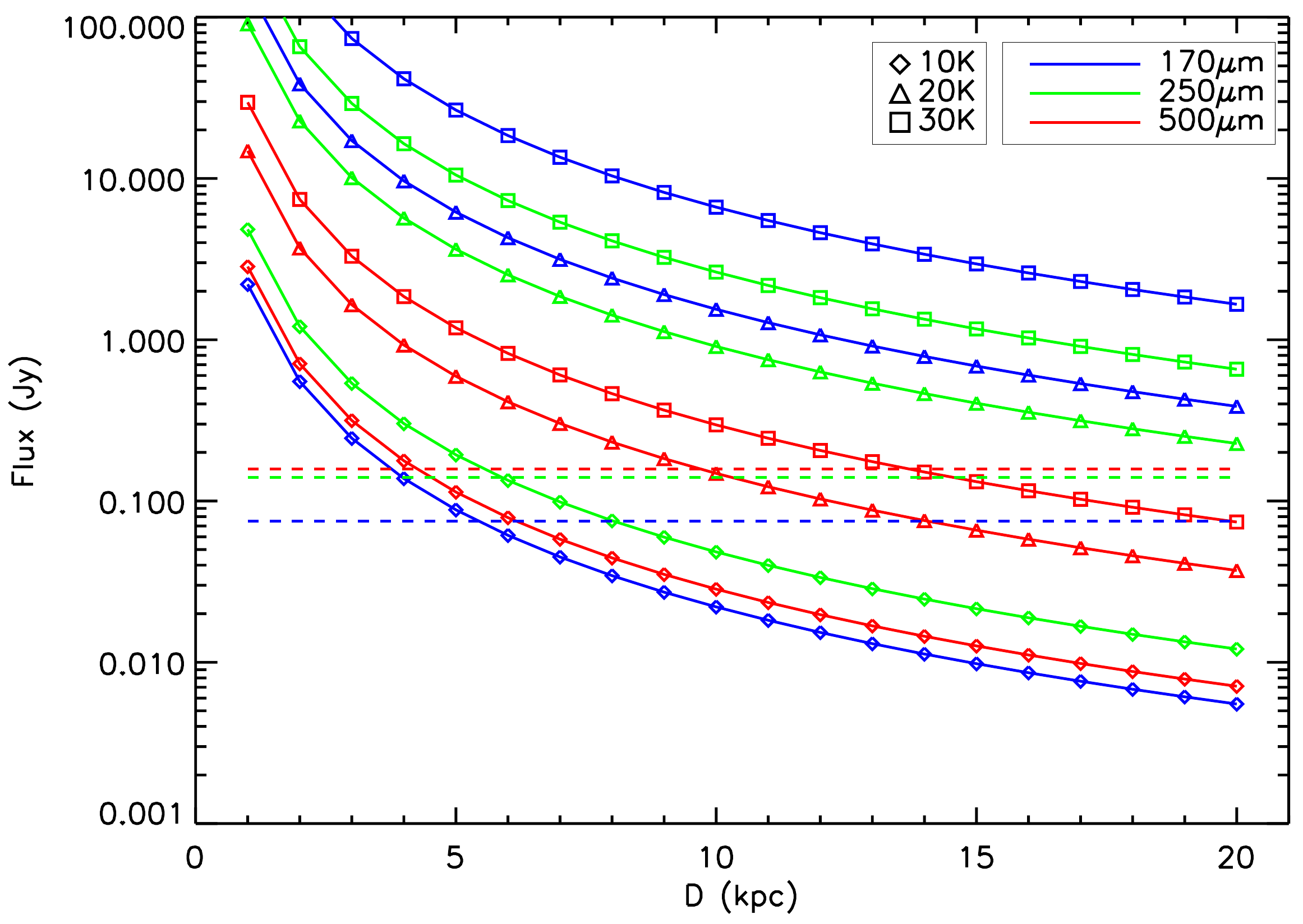}}
\caption{Flux expected in the 170, 250 and 500\um\ Herschel bands (color coded) from a 20\msun\ core (dust+gas) for different temperatures (different symbols), as a function of distance in kpc. The dashed lines represent the confusion noise expected in the three bands.} 
\label{masslimit}
\end{figure}

The 1-$\sigma$ sensitivities provided by HSPOT, for a single
Astronomical Observation Request (AOR), are 17.6 and 26.8\,mJy in the
two PACS 70 and 170\um\ bands, 
and 12.8, 17.6 and 14.9\,mJy for the  SPIRE bands; co-addition
of the orthogonal scanning patterns will provide $\sqrt{2}$ better
figures. These sensitivities result from the adopted scanning strategy
designed to maximize redundancy and map fidelity especially for large scale
diffuse structures. However, the limiting factor for the detectability of sources and clouds will likely be cirrus confusion. Estimates based on recent BLAST measurements (\citealt{roy10}) suggest values of the order of 75, 140, and 160 mJy in the 170, 250 and 500\um\ Herschel bands for a representative region of the Galactic Plane at $l$=45\adeg; these values are greater than the detector sensitivities. Fig.~\ref{masslimit} shows the expected flux from a 20\msun\ envelope as a function of distance (in kpc) in each of three bands above mentioned, and for three different dust temperatures; we adopted $\beta$=2 and the dust opacity from \citet{preib93}. The horizontal dashed lines (color-coded with wavelength) represent the predicted confusion noises based on BLAST images. Fig. \ref{masslimit} shows that we will detect the representative 20\msun\ core everywhere in the Galaxy except for very cold dust (T$\leq$10K), for which detectability is predicted to be limited within a distance of about 5 kpc. We may then conclude that cirrus confusion is not going to be a problem for the investigations of the intermediate and high-mass star formation studies which are the "core science" of this project (see \S\ref{timeline} and \ref{global}).

\subsection{Detection of extended structures}

\begin{figure}[t!]
\includegraphics[width=8cm]{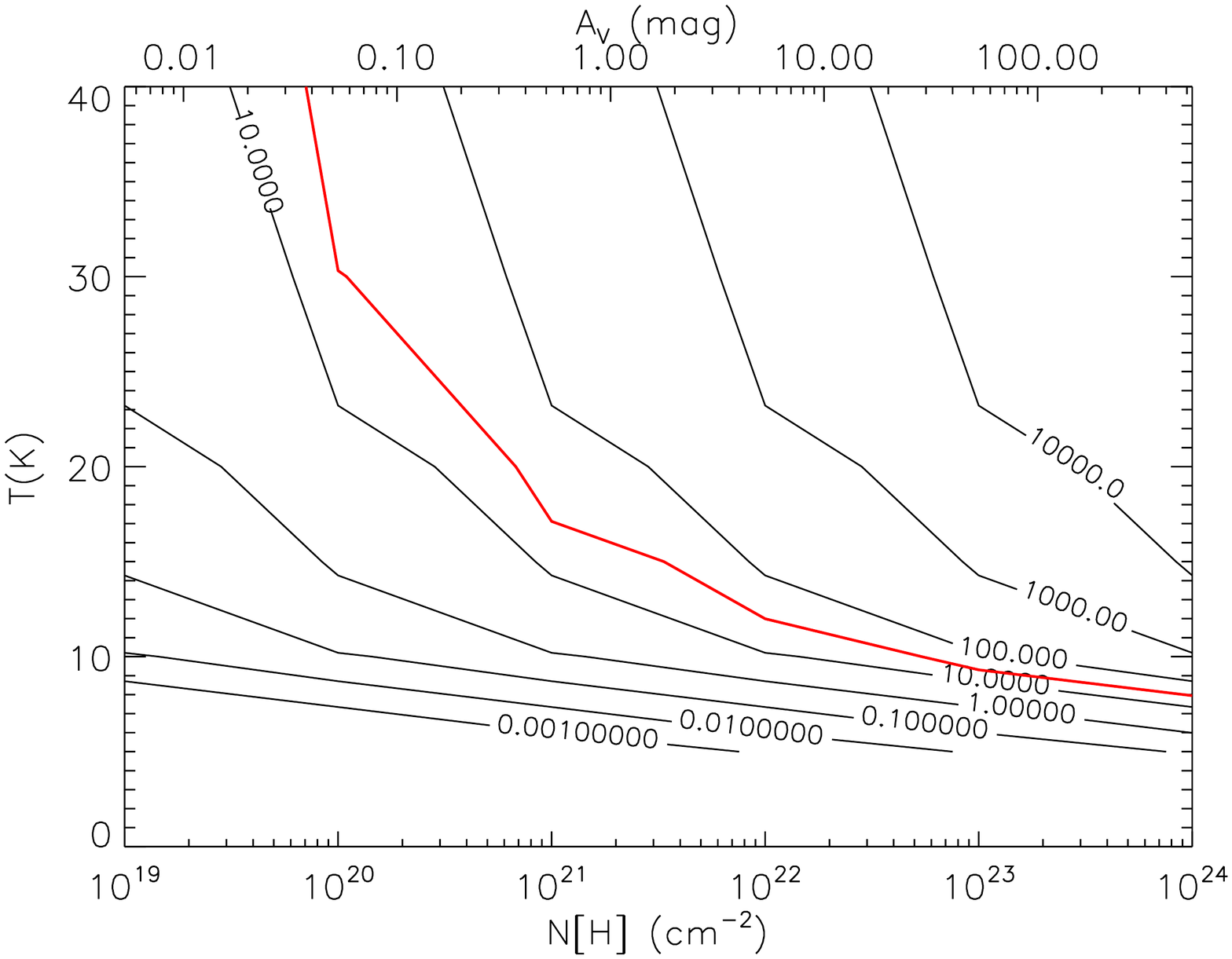} \\
\includegraphics[width=8cm]{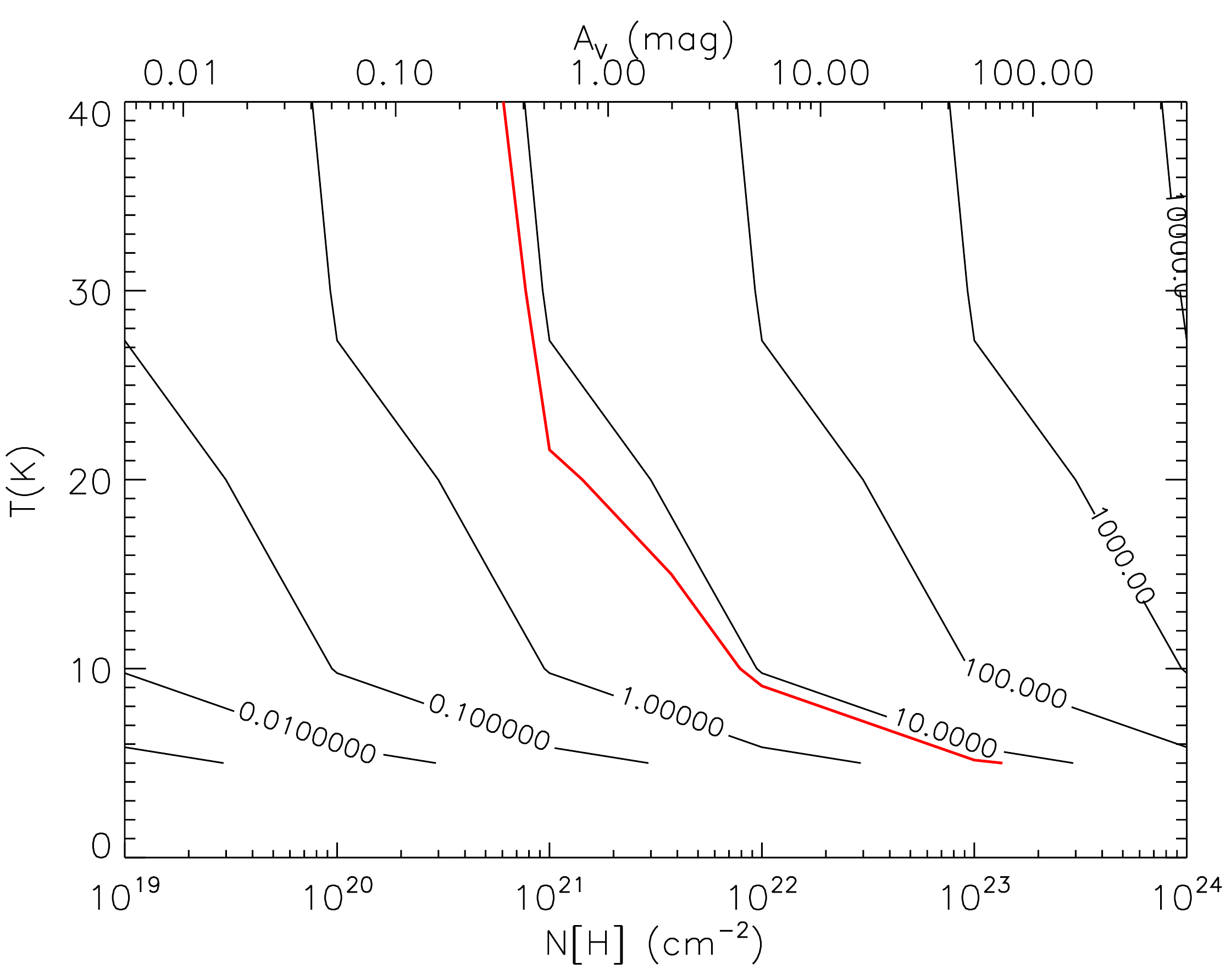} 
\caption{Brightness (in MJy/sr) of optically thin dust as a function
  of temperature and column density at 170\um\ (top) and
  500\um\ (bottom). The red lines indicate the expected confusion  noise.} 
\label{diffuse}
\end{figure}

The diffraction-limited instrument beams at all wavelengths can be used to translate the confusion noises reported in the previous section into brightness units to investigate the detectability limits expected for diffuse extended structures. Calculations of the expected brightness levels
from optically thin dust as a function of temperature and H column
density (assuming gas/dust=100)  are reported in Fig. \ref{diffuse}a,b for $\lambda$=170 and 500\um\ respectively, where the expected confusion noise levels are also reported in red lines.

The figures show that with a typical confusion noise of about 
10 MJy/sr at 500\um\ it will be possible to detect most Infrared Dark Clouds (IRDCs, see \S\ref{cold_dust}), where column densities are in the range
10$^{23}$-10$^{25}$cm$^{-2}$. Less dense clouds
with column densities 
of the order of few 10$^{21}$cm$^{-2}$  should
be easily detectable at levels of tens of MJy/sr at 
170\um\ at temperatures as low as T$\sim$20K. 
The situation may be less simple in the regions closer to the Galactic center. However, our broad spectral coverage provides an important
advantage for measuring the temperature accurately, and for isolating
structures and sources with temperature different from the standard
diffuse ISM cirrus ($\sim$20K).  Besides, all of the above is based on extrapolated estimates and we will provide the definitive measurement of the cirrus confusion at Herschel wavelengths and resolution. 


\subsection{Checks on photometric calibration}

Virtually every block of the Hi-GAL survey will contain secondary
calibrators, either stars or other well known objects, ensuring 
accurate checks of the flux calibration of \higal\ data. Our baseline calibration targets will be the 400 stars
used as calibrators for the Spitzer GLIMPSE-I/II surveys. These are
mostly A0-5V or K0-M0III stars, although they also include $\sim$60
calibrators of hot or warm dwarfs (B-G). In order to remove sources
with non-photospheric FIR emission we will make predictions for the
24\um\ fluxes and then test for excesses in the MIPSGAL data first.
Once anomalous sources are excluded we can extrapolate to the FIR and
create an initial set of calibrators.  

We will also be able to obtain a reliable calibration for extended
sources, which is one of the more difficult parts of the nominal
instrumental calibration activity. As part of the calibration scheme
we will compare fluxes in the SPIRE 500\um\ band with fluxes from the
same band of Planck-HFI in suitable locations\footnote{many of the Hi-GAL Co-Is are also Planck Consortium members}.

\section{\higal\ Key Science}

\subsection{The Distribution of the ISM Temperature and the Intensity
  of the Interstellar Radiation Field} 

At near-infrared wavelengths, the emission from dust is produced 
by small particles whose abundance varies significantly, being 
strongly depleted by coagulation processes in the dense ISM (e.g. 
\citealt{ber93,abe94,ste03}). Far-infrared (FIR) emission is produced
by larger grains which are more stable and dominate the total dust
mass and trace all 
phases of the ISM.  The ISM dust spectrum peaks in the FIR where the
Galaxy is transparent.  FIR emission is therefore a reliable tracer of
the overall ISM column density structure in our Galaxy. Other 
phase-independent tracers include dust absorption and
gamma-ray production, where however the former can be used to sample only the
nearest 1\,kpc, and $\gamma$-ray surveys currently lack sensitivity
and angular resolution.

Variations in the FIR emissivity (the ratio of surface brightness to
column density) are dominated by the non-linear effects of dust temperature
through the Planck function. Fortunately, the shape of the dust SED as
measured by PACS and SPIRE will be most sensitive to temperature
variations as the spectral bands sample the peak of the Big Grain
emission and the contribution of Very Small Grains can be estimated
from the \higal\ data at 70\micron\ and MIPSGAL at 24\micron.  The dust
temperature ($T_d$) and its spatial variations will therefore be
measured precisely. This important parameter can be used, in conjunction with complementary data from Planck, HI, CO, H$\alpha$ and $\gamma$-ray surveys , to estimate 
the strength and spectral shape of the InterStellar Radiation Field (ISRF),
which is set by the stellar content in a given region. So far, the dust temperature 
in the Galactic Plane has been mapped over limited regions using IRAS (\citealt{kim99}, \citealt{dougtay07}, and over the entire Plane at a resolution of 40\arcmin\ (\citealt{lag98}) using DIRBE. \higal\
will improve with respect to the latter by a factor of about 100 in linear scales.  It will
trace the local radiation field on scales relevant to star formation,
and provide mass estimates even at large distances. In the case of the dense medium, determining the 3-D distribution of  
the ISRF
strength and spectral shape in a given cloud will require radiative  
transfer modeling.
This is possible, even for complex geometries, using Monte-Carlo codes  
(e.g. \citealt{juv03}).
Using such codes, the equilibrium dust temperature $T_d$ and the dust  
emission can be predicted at
any 3D location in the cloud. Integration along the line-of-sight in  
turn allows to predict 2D emission maps.

\subsection{Molecular cloud formation}
\label{cold_dust}

\begin{figure*}[t!]
\resizebox{\hsize}{!}{\includegraphics[width=7.cm]{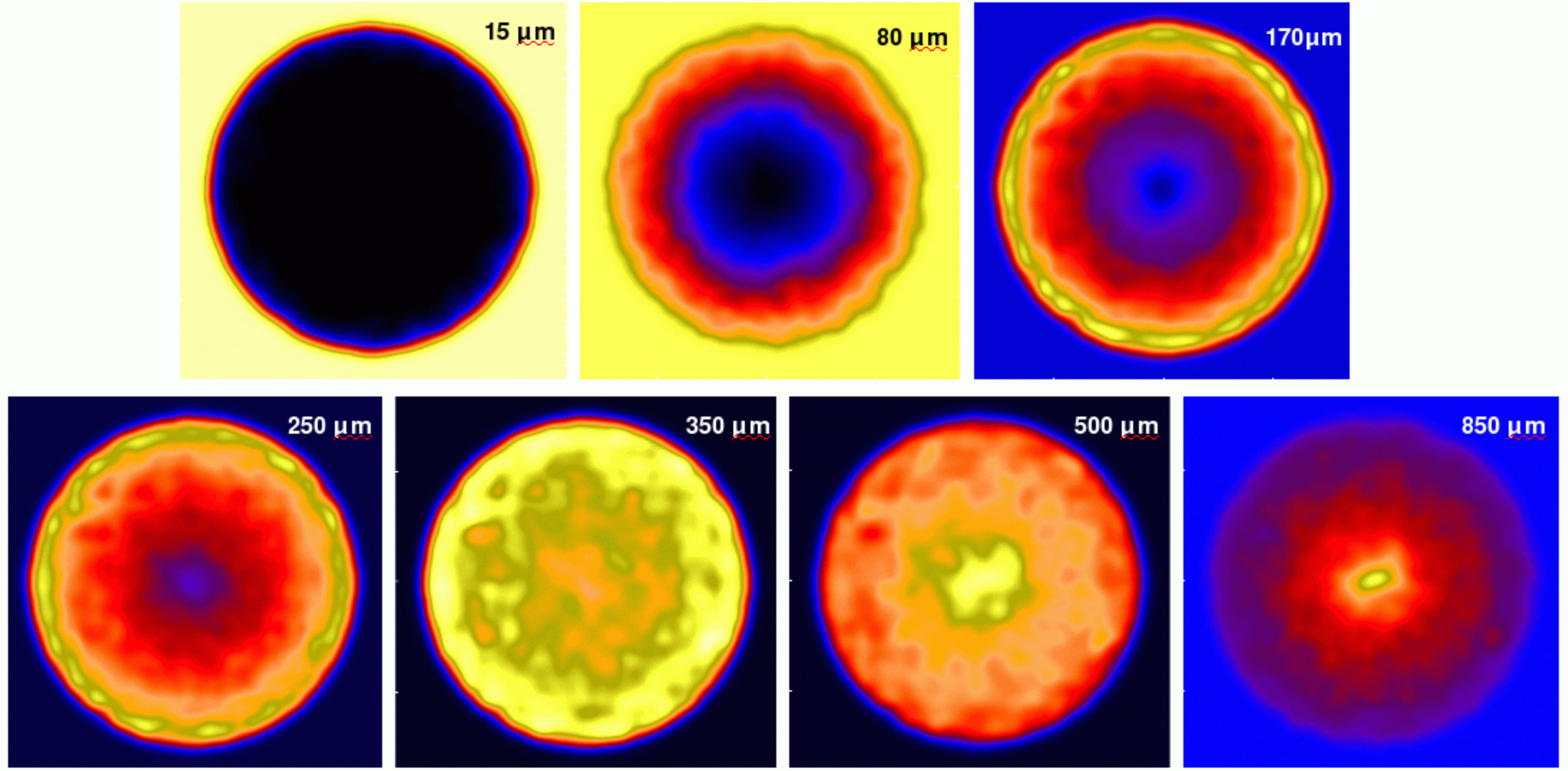}}
\caption{Model of a typical IRDC after \citet{sta04} at seven different wavelengths 
illustrates the importance of the Herschel wavelength coverage and resolution 
(at 2\,kpc the linear extent is 5\arcmin) to obtain measurements of
IRDCs in emission. } 
\label{irdc}
\end{figure*}

About a quarter of the mass in the ISM is in
molecular form \citep{blitz97} and most of that material resides in giant
molecular clouds (GMCs). Since GMCs are also the dominant sites of
star formation, understanding their origins and evolution is essential
to our understanding of the Galactic environment. 

In combination with molecular line surveys, \higal\ will provide the data needed to derive basic physical properties of GMCs. 
We will detect and characterize cold structures in the
inner GP and classify them based on star formation activity. Detection
statistics for clouds with different temperatures and degree of star
formation activity will provide the fraction of quiescent \textit{vs}
star-forming clouds.
It will thus be possible to constrain the properties and lifetimes of GMCs in
our Galaxy and to compare with the predictions of fast evolution of molecular clouds (\citealt{har01}) or a more
traditional slow evolution of star formation in our Galaxy (\citealt{shu87}).
Variations with Galactocentric radius will
determine if the \textit{slow/fast} scenarios are mutually exclusive
or reflect different initial/environmental conditions. A large-area survey like \higal\ will provide the needed statistical 
significance in all mass bins, especially at the high-mass
end,  and in a variety of Galactic environments.  

Direct detection of cold (i.e. T$<$20 K) dust which could be the  quiescent
counterparts to GMCs, is difficult
(\citealt{sod94,rea95,lag98}) either because of insufficient wavelength
coverage (e.g. IRAS) or inadequate spatial resolution (DIRBE, FIRAS). CO
observations are problematic due to molecular freeze-out onto grains
(\citealt{flo05}), or photo-chemical effects in low-metallicity
environments (\citealt{bot07}). The recent detection of very cold clumps
in the GP with Archeops (\citealt{des08}), confirms the FIR and submm
continuum as the best tool to trace cold ISM
components. Notable examples are Infrared Dark Clouds (IRDCs) and \hi\
Self-Absorption (HISA) clouds.

IRDCs are structures initially discovered as extinction features against the
bright mid-IR Galactic background, and soon verified to exhibit properties similar to molecular clouds. Their properties ($n>10^5$\,cm$^{-3}$,
N$_H \sim 10^{22}-10^{24}$ cm$^{-2}$ and $T<25$\,K -
\citealt{ega98,car00}, $R \sim 5$\,pc and $M\sim10^3$\,M\subsun
\citealt{sim06,rat06}) suggest that they are the precursors of
cluster-forming molecular clumps like Orion. Thus, IRDCs are ideal for
the study of the pristine, undisturbed physical conditions that may
produce massive stars and clusters. IRDCs have only been detected
against the bright Galactic mid-IR background (mostly for $|l|\leq
30^{\circ}$); their true Galactic distribution is unknown. Modeling of
IRDCs in the IR and submm (Fig. \ref{irdc}, \citealt{sta04}) proves
Herschel's unique ability  to detect them and to measure their SEDs. IRDCs,
with $\tau_{200\mu m}\ge 1$, are not detectable by either IRAS or
Spitzer. \higal\ will provide a definitive inventory of cold dust and
potential sites of massive star cluster formation everywhere in the
inner GP.

HISAs are traced by cold \hi\ gas seen in absorption against a background of warm \hi\ emission (\citealt{gib00, gib05,gol05}); they may provide additional clues to the formation of molecular clouds. When compared with molecular tracers, these cold (relative to the ambient neutral medium) and relatively quiescent ($\Delta v \sim 1-3$\,kms) clouds show a wide range of \hi /H$_2$ ratios (\citealt{li03,kla05}) which suggests that they might be \hi $\rightarrow$H$_2$ conversion sites. Detailed studies of their FIR$\rightarrow$mm SED shapes can help clarify this issue, providing evidence for the grain types necessary for a reasonable \hi $\rightarrow$H$_2$ formation timescale (\citealt{gol05}). 

\subsection{Timeline of high-mass star formation}
\label{timeline}

\begin{figure*}[t!]
\resizebox{\hsize}{!}{\includegraphics[width=8.5cm]{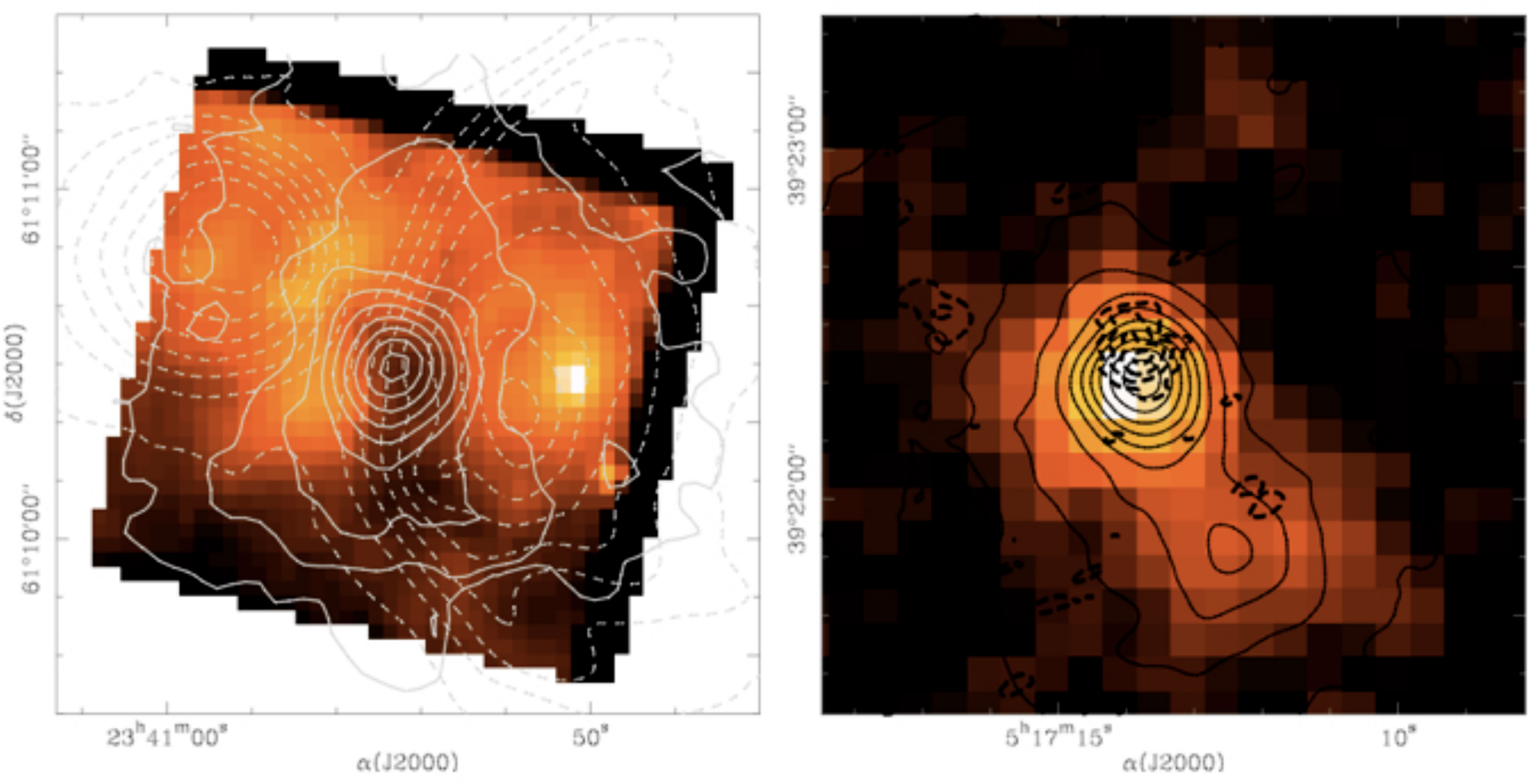}}
\caption{Mid-IR 21\um\ MSX images for a candidate precursor of
  a Hot Core  
(IRAS23385+6053, left), and a UC\hii\ (IRAS05137+3919, right). Full and dashed 
contours represent the millimeter (cold dust) and radio (jet)
continuum emission (from \citet{mol08}). \higal\ will complement these with similar
resolution FIR data.} 
\label{molsources}
\end{figure*}

The paradigm for the formation of solar-type stars via accretion through
a circumstellar disk (\citealt{shu87}) predicts an evolution from cores
to protostars and, finally, pre-main sequence stars that is well matched
with distinctive characteristics of their SEDs (\citealt{lad84,and93}).
The empirical classification of the SED of low mass YSOs has thus been
used as a powerful tool to constrain theoretical models.

Higher mass stars reach the conditions for H-burning faster than the
time required to assemble them, so that winds and
radiative feedback 
will strongly influence accretion and may limit the
final mass of the star (\citealt{zin07}). However, since massive stars
exist, several theories have been proposed to solve this puzzle
including accretion from a disk, a very high-pressure ambient medium,
"flashlight effect" (\citealt{yor02,mck03}), competitive accretion, or
coalescence (e.g. \citealt{sta00}). Application of SED-based
classification tools, and evolutionary diagnostics like the M$_{env}$-L$_{bol}$ diagram which relates the bolometric luminosity of a YSO to the mass of its envelope   \citep{mol08}, to a large sample of luminous protostar candidates
in the GP will define a timeline for the various phases of massive star
formation that will constrain the theories and lead to new estimates of
the SFR. Clearly the source distance is the crucial parameter here; we are collecting  the information from the major molecular line surveys over the inner Galactic Plane, while planning to undertake additional surveying activities at a variety of facilities in several high-density tracers to get additional data (see \S\ref{othersurveys} for more detail).

An evolutionary sequence for massive YSOs has been proposed (cold massive 
cloud core; Hot Molecular Core with outflow; IR-bright massive YSO; 
ultracompact (UC) \hii\ region, e.g. \citealt{eva02,kur00}) but it is 
qualitative and based on small and possibly incomplete samples.  Samples 
of bright and massive YSOs
(\citealt{mol96,sri02,hoa04}) are IRAS or MSX selected and tend to
suffer from age biases and confusion which prevent firm quantitative
conclusions. A phase of intense and accelerating accretion prior to
H-burning ignition, that may be observable (e.g. \citealt{mol98}) in the form
of dense condensations devoid of IR as well as radio continuum emission,
seems confirmed by recent large mm surveys (\citealt{bel06,hil05}).
Millimeter continuum alone, however, cannot distinguish between
pre-collapse condensations and rapidly accreting cores; \higal\ will use
the full potential of Herschel wavelength coverage and spatial
resolution to trace the SED peak of dust envelopes in all phases, from
massive pre-stellar condensations to UC\hii\ regions. An angular resolution of 30\arcsec\ or less,
typical of Hi-GAL, has been proven (e.g. \citealt{mol08}) to be the key
to building accurate  
SEDs, deriving reliable luminosities for massive YSOs, and distinguishing embedded UC\hii s strongly 
emitting in the Mid-IR and radio from pre-UC\hii s (Fig. \ref{molsources}). 

The abundance of high-mass YSOs per mass bin in the various evolutionary
phases will provide an estimate of the duration of each phase. This timeline can be
directly compared with the predictions of various models, and together
with the YSO mass function, will be used to infer the SFR. As an
example, using current estimates for SFR and IMF
(\citealt{mck97,kro01}), and a 10$^5$ yr period to assemble a massive
star (\citealt{mck03}) , we expect $\ge$ 1000 FIR/sub-mm objects with
M$>$15M\subsun\ in the Galaxy, and only $\sim$180 objects with
M$>$50M\subsun. Recent millimeter surveys in the Cyg X region
(\citealt{mot07}) confirm the rarity of such massive precursors and
strengthen the need for a systematic unbiased search. 

\subsection{Bridging the Gap between Global and local star formation}
\label{global}

Galactic phenomenology currently invokes an indeterminate mixture of
spontaneous and triggered star formation. Triggering agents include
radiation pressure from OB stars (\citealt{sug89}), compression by
expanding \hii\ regions (\citealt{elm77,deh05}), or fragmentation of
supershells by multiple supernovae in OB associations (\citealt{mcc87}).
On larger scales we still do not know if spiral density waves actively
induce star formation (e.g., \citealt{elm02}) or simply assemble
star-forming regions, with local feedback and triggering becoming more
important within the arms (\citealt{sle96}). The mean SFE of a galaxy
can increase (up to 50 times) in starbursts (\citealt{san91}) and galaxy
mergers due to strong feedback effects, a process observed in miniature
in Galactic star-forming regions (e.g. \citealt{moo07}). Whether the IMF
depends on local triggering and other environmental factors is unclear.

\higal\ will enable quantitative analysis based on basic observables -
the luminosity functions of YSOs, the mass function of dense
star-forming structures and quiescent clouds. Hi-GAL will provide the
essential context of high-mass star formation, as it relates to molecular
gas, \hi\ gas, stars, \hii\ regions, OB associations, SNRs and spiral
arms. Theoretical models and numerical simulations will be tested in
multiple ways. We will discover whether a local triggering agent is
necessary  for high-mass star formation or if a spiral arm is
sufficient, clarifying the differences between spontaneous and
triggered star formation. We will quantify the relationship between
the interaction strength (estimated using available data from
ancillary surveys) and the resulting increase in SFE above the
spontaneous rate. By locally relating the SFR to the properties of the
ISM we will probe star formation thresholds as a function of
environment and spatial scale, and possibly unveil the mechanism
giving rise to global Schmidt-like scaling laws. We will determine the
dominant physical process underlying triggering.

\subsection{Serendipitous Science}

The 5-band FIR images and source catalogues
provided by \higal\ will allow research in many fields that can only be
partially anticipated. A detailed description of the specific outcomes
in all these fields goes beyond the scope of the present paper. These
aspects will of course be the object of dedicated publications to be
released in due time. In the following we list some of the possible by
products of \higal\ :
\begin{list}{\labelitemi}{\leftmargin=1em}
\item An input catalogue for ALMA: we expect to detect some 200-400
  objects per tile, most of which will mark very cold objects to be
  studied at all possible wavelengths;
\item complete characterization of the Galactic foreground in the
Far-IR and submillimeter, critical for the correct interpretation and
modeling of cosmological backgrounds; 
\item Dust formation and destruction in supernovae
remnants
\item Debris dust disks around main sequence stars,
with unbiased statistics on frequency and mass as a function of star
age
\item Evolution of dust properties, especially around AGB stars, the
  factories of cosmic dust; 
\item Detection of detached dust shells around first
ascent giant stars to investigate missing mass in AGB envelopes;
\item Detection of multiple shells around AGB stars, post-AGB
objects and planetary nebulae, as well as around various classes of
interacting binaries; 
\item Detection of ejecta shells and
swept-up ISM bubbles around massive stars, providing a complete census
of WR and LBV stars; 
\item Extinction maps to aid in correcting
Near-IR galactic star counts; 
\item Detection of Solar system
objects via comparison of cross-linked rasters: in particular the
detection of asteroids will be very interesting for studies of the
albedo
\item Nearby Low-Mass SFRs in the GP: Herschel will detect
  many nearby star forming regions and individual YSOs.
\end{list}

\section{\higal\ and its place in the context of the Multi-Wavelength
  Milky Way}
\label{othersurveys}

\begin{figure}[h!]
\resizebox{\hsize}{!}{\includegraphics[width=8.3cm]{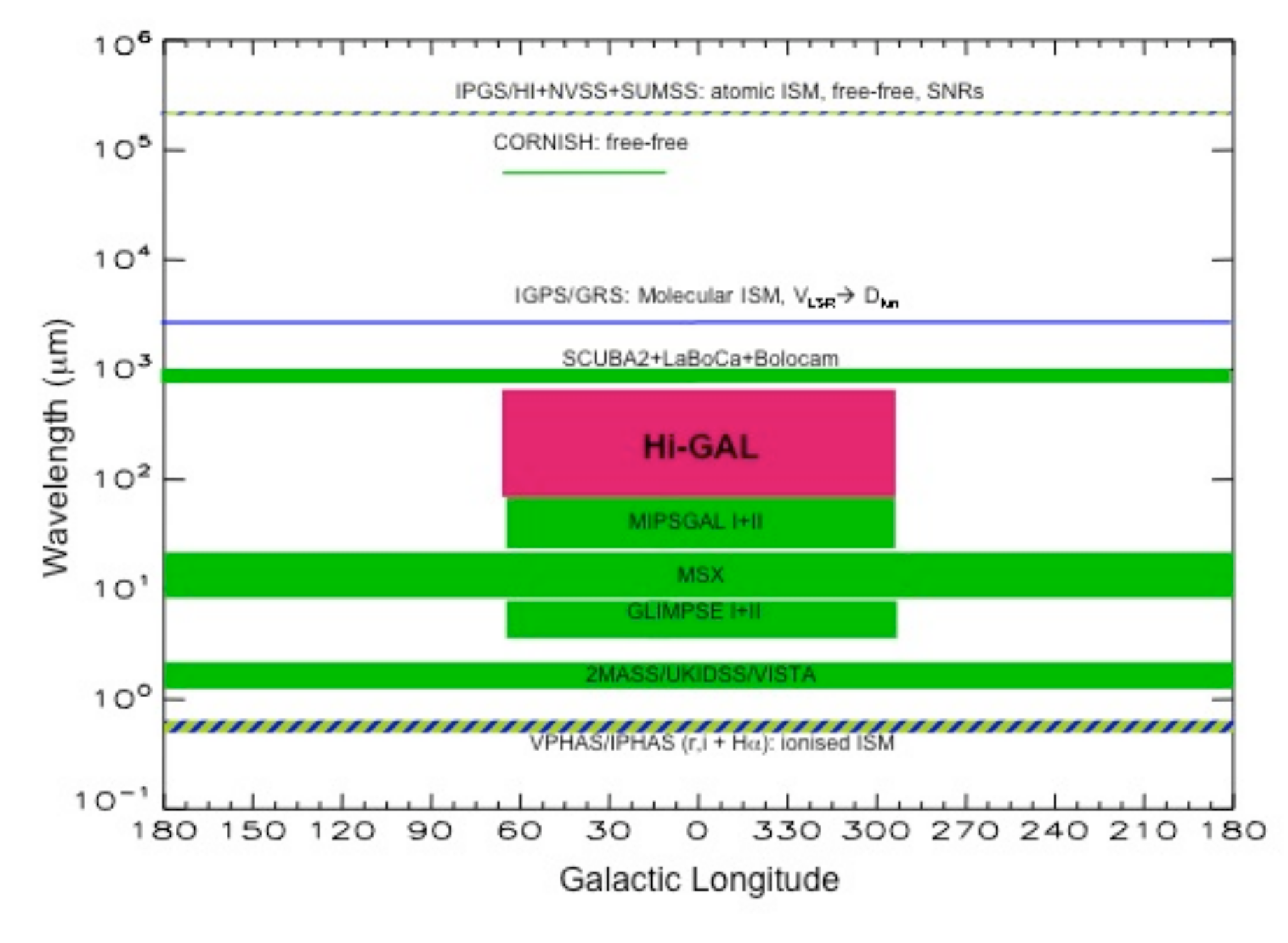}}
\caption{Wavelength-longitude coverage plot for photometric (green) and
spectroscopic (blue) existing or planned GP surveys for the next decade.
Hi-GAL (red) fills the critical gap
where the ISM dust emission peaks, between MIPSGAL and the
SCUBA2/LaBoCa/Bolocam surveys.}
\label{higal_context}
\end{figure}

PACS and SPIRE are unique in tracing the peak of the Spectral
Energy Distribution (SED) of cold dust and, hence, temperatures and
luminosities of both star-forming complexes and the ISM, at 
resolutions unmatched by any previous instruments.  However, the full 
potential of the \higal\ survey will be realised in the context of 
the other unbiased Galactic Plane surveys which are shaping our 
understanding of the Galactic ecosystem. A suite of surveys in the 
mid- and far-infrared continuum,
ISOGAL (\citealt{omo03}), MSX (\citealt{pri01}), GLIMPSE
(\citealt{ben03}), MIPSGAL (\citealt{car05}), IRAS (\citealt{neu84})
and Akari (\citealt{mur07}), has been
and will be complemented by surveys in the submillimeter and
millimeter spectral range including the BGPS survey with Bolocam at 1.1mm
(\citealt{ros09}), the ATLASGAL survey currently underway 
with the LaBoCa camera at APEX (\citealt{schu09}), and the SCUBA2 JPS 
survey beginning in 2010 (Fig.~\ref{higal_context}).

BLAST and AKARI have wavelength coverage and resolution not too different form PACS and SPIRE. However, BLAST has been used to map limited portions of the GP while results from AKARI photometric imaging of the GP are not yet found in the literature.

\higal\ will be the scientific keystone of this suite of
surveys, completing the continuous coverage of the dust continuum 
over three orders of magnitude in wavelength at sub-30\arcsec\ resolution, 
and allowing the measurement of dust temperatures and luminosity over 
the inner Galactic Plane. 

An extensive plan for radio spectroscopic cross-correlation and follow
up of the \higal\ survey has been devised. The top priority is for radio
spectroscopic observations to obtain distance estimates for detected
sources and structures. CO and $^{13}$CO data at sub-arcminute spatial
resolution are available via the IGPS for the first Galactic Quadrant (Q{\sc I});
lower resolution CO data are available from the NANTEN survey (Q{\sc
IV}). Should CO
resolution not be sufficient, e.g. in case of very cold cores with CO
depletion, we plan extensive follow-up programs (e.g. N$_2$H$^+$,
NH$_3$) at the Mopra 22m antenna (Australia) for Q{\sc
IV}, and at Green Bank, Effelsberg, Medicina and Onsala for Q{\sc
  I}. HI is also available for Q{\sc I} and {\sc IV} at 
sufficient resolution from the IGPS to help resolve the distance
ambiguity in the inner Galaxy \citealt{bus06}. H$\alpha$ VPHAS+IPHAS surveys will be
also used.

In addition to the kinematic distance estimates, we plan to use NANTEN2,
Mopra and APEX for detailed multiline studies of evolution-sensitive
chemical tracers (N$_2$H$^+$, NH$_3$, CS, HCN, CH$_3$OH, CH$_3$CN, etc.)
toward clouds and objects discovered by \higal. \higal\ catalogues
will be 
the primary source of major future high spatial resolution follow ups in
the sub-mm with ALMA; in the meantime such programs will be attempted
using the SMA interferometer through Legacy class proposals.

\section{Data Processing and Products}

It is relatively easy to translate our scientific goals into a clear set
of requirements on the data processing: we require that the dust continuum
emission be detectable, and accurately measurable, at all bands over the
broadest range in signal levels (down to the confusion limit) and
spatial scales. The observing strategy is carefully designed to that
effect, but ensuring that this information is properly extracted from
the data stream over a 240 sq.deg. area is a formidable challenge. We
will use the Herschel Interactive Processing Environment (HIPE) for all
those processing steps dealing with fundamental instrumental calibration
and issues, but we anticipate areas where a dedicated set of specialised
tools can take advantage of the homogeneous observing strategy and
deliver higher quality results compared to the standard pipeline
products; pointing refining, map-making, source extraction and
photometry are examples.

We will make available a set of data products which will include maps
and compact source catalogues at the five \higal\ wavelengths. These
products will be made available after the end of observations (EoO) for
the entire survey via incremental releases.

Improved reprocessed maps and source catalogues will be subsequently
made public, together with a first release for an extended source
catalogue. All public deliveries will be accompanied by an Explanatory
Supplement.

In addition to this minimum set of products, quite standard for any
large-scale survey like \higal, we plan to make available to the
community a set of scientific value-added products which will be created
during our scientific analysis, including band-merged catalogues
integrated with data from continuum surveys at adjacent wavelength,
color maps, source-subtracted maps. The public access to this final set
of products is foreseen for EoO+42 months.

\section{Conclusions}

\higal\ is an Open Time Key Project to be performed with the 3.5\,m orbiting
Herschel telescope, to map photometrically the inner Milky Way ($\mid
l \mid < 60^{\circ}, \mid
b \mid < 1^{\circ}$) in five wavebands between 70$\mu$m and 500$\mu$m
simultaneously, using $\sim$350 hours of observing time. The unique
combination of survey speed, high sensitivity, high spatial
resolution and wavelength coverage (across the peak of the dust
emission) make \higal\ the first dedicated project to study the early
phases of GMC- and high-mass star formation in the Galaxy, with a
legacy value similar to the IRAS mission some 20 years ago. The
outcomes of \higal\ will consist of source lists and images to be
released in due course after EoO.

\acknowledgments

We are grateful to all the people who made the building and launch of
Herschel such a success. In particular ESA and the Herschel Project
Scientist G. Pilbratt, and the instrument teams of PACS and SPIRE
magnificently led by A. Poglitsch (MPE, Garching) and M. Griffin
(Univ. of Cardiff).

{\it Facilities:} \facility{Herschel}, \facility{SPIRE}, \facility{PACS}.

\bibliography{herschel_bib}
\bibliographystyle{aa}

\end{document}